%% file: main.tex
\documentclass{IEEEtran}

\usepackage{cite}

\ifCLASSINFOpdf
   \usepackage[pdftex]{graphicx}
\else
   \usepackage[dvips]{graphicx}
\fi
\usepackage{standalone}
\usepackage{pgf, tikz}
\usepackage{pgfplots}
\usepackage{pgfplotstable}
\usepackage{verbatim}
\usetikzlibrary{arrows}
\usetikzlibrary{decorations.text}
\usetikzlibrary{shapes.geometric}
\usetikzlibrary{shapes.arrows}
\usetikzlibrary{arrows}
\usetikzlibrary{shapes}
\usetikzlibrary{positioning}
\usetikzlibrary{shadows}
\usetikzlibrary{patterns}
\pgfplotsset{plot coordinates/math parser=false}
\pgfplotsset{compat=newest}
\pgfplotstableset{use comma,1000 sep=\,}
\pgfkeys{/pgf/number format/.cd,fixed,precision=2} 
\usepgfplotslibrary{groupplots}

\usepackage{amsmath,amssymb,amsfonts}
\usepackage{mathtools}
\usepackage[nohyperlinks, nolist]{acronym}
\usepackage{algorithm}
\usepackage{algpseudocode}

\ifCLASSOPTIONcompsoc
  \usepackage[caption=false,font=normalsize,labelfont=sf,textfont=sf]{subfig}
\else
  \usepackage[caption=false,font=footnotesize]{subfig}
\fi

\begin{document}
\bibliographystyle{IEEEtran}

\title{Representation of Distribution Grid Expansion Costs in Power System Planning}

\author{
\IEEEauthorblockN{Luis Böttcher, Christian Fröhlich, Steffen Kortmann, Simon Braun, Julian Saat, Andreas Ulbig}
\IEEEauthorblockA{\textit{Institute of High Voltage Equipment and Grids, Digitalization and Energy Economics (IAEW) at RWTH Aachen University} \\
Aachen, Germany\\
l.boettcher@iaew.rwth-aachen.de}
}

\maketitle

\IEEEpubidadjcol

\begin{acronym}[TDMA]
    \acro{AC}{alternating current}
	\acro{ACPF}{alternating current power flow}
	\acro{ADN}{active distribution network}
	\acro{CAPEX}{capital expenditures}
    \acro{CEP}{capacity expansion problem}
	\acro{CHP}{combined heat and power plant}
	\acro{CLC}{CORINE land cover}
	\acro{COP}{coefficient of performance}
	\acro{DAC}{direct air capture}
	\acro{DER}{distributed energy resources}
	\acro{DG}{distribution grid}
	\acro{DHS}{district heating system}
	\acro{DSM}{demand side management}
	\acro{DSO}{distribution system operator}
	\acro{EAA}{equivalent annual annuity cost}
	\acro{ENTSO-E}{European Network of Transmission System Operators}
	\acro{ENTSOG}{European Network of Transmission System Operators for Gas}
	\acro{ESS}{energy storage system}
	\acro{EV}{electric vehicles}
	\acro{FACTS}{flexible AC transmission systems}
	\acro{FOM}{fixed operation and maintenance cost}
	\acro{FOR}{Feasible Operation Region}
    \acro{FPR}{Feasible Planning Region}
	\acro{FPU}{flexibility providing units}
	\acro{GDP}{gross domestic product}
	\acro{HP}{heat pump}
	\acro{ICT}{information and communication technology}
	\acro{IPF}{interconnection power flow}
	\acro{KCL}{Kirchhoff's current law}
	\acro{LCT}{low-carbon technologies}
	\acro{LP}{linear program}
	\acro{MES}{multi-energy system}
	\acro{MILP}{mixed-integer linear program}
	\acro{MESOM}{multi-energy system optimization model}
	\acro{OP}{optimization problem}
	\acro{OPEX}{operational expenditures}
	\acro{OPF}{optimal power flow}
    \acro{OCGT}{open-cycle gas turbine}
    \acro{CCGT}{combined-cycle gas turbine}
	\acro{PF}{power flow}
	\acro{PtX}{power-to-x}
	\acro{PtG}{power-to-gas}
	\acro{PtH}{power-to-heat}
	\acro{PtH2}{power-to-hydrogen}
    \acro{PV}{photovoltaic}
	\acro{PyPSA}{Python for Power System Analysis}
	\acro{RES}{renewable energy sources}
	\acro{TES}{thermal energy storage}
	\acro{TSO}{transmission system operator}
    \acro{TG}{transmission grid}
	\acro{V2G}{vehicle-to-grid}
	\acro{VOM}{variable operation cost}
    \acro{TYNDP}{Ten-Year Network Development Plan}
    \acro{MV}{medium-voltage}
    \acro{LV}{low-voltage}
\end{acronym}
\input{00_abstract}


\input{01_introduction}

\input{02_methodology}

\input{02_modelling}

\input{03_results}

\input{04_conclusion}

\bibliography{bibliography}

\end{document}

%% file: 00_abstract.tex
\begin{abstract}
The shift towards clean energy brings about notable transformations to the energy system. In order to optimally plan a future energy system, it is necessary to consider the influence of several sectors as well as the interaction of the transmission grid and distribution grid. The concept of Feasible Operation Region (FOR) is a detailed approach to representing the operational dependencies between the transmission and distribution grid. However, in previous planning procedures, only a simplified expansion of the distribution grids can be taken into account. With the method presented in this paper, a Feasible Planning Region (FPR) is developed, which represents the operational boundaries of the distribution grids for several expansion stages and thus represents an admissible solution space for the planning of distribution grids in systemic planning approaches. It hence enables a more detailed representation of the necessary distribution grid expansion for the integration of distributed technologies in an optimized energy system of the future. In this paper, we present the method by which the FPR is formed and its integration into an energy system planning formulation. In the results, the FPR is presented for different voltage levels, and its use in power system planning is demonstrated.

\end{abstract}

%% file: 01_introduction.tex
\section{Introduction}
With the advance of climate change and the accompanying technological change in energy systems, a trend towards decentralized generation, e.g. through photovoltaic and wind power, can already be observed today. In addition, changes in the heating and mobility sector towards electric heat pumps and electromobility are to be expected in the future, which will mainly be connected to the distribution grids. To adequately consider this change in the supply task in the planning of future energy systems, there are methods for energy system planning. There are recognized methods for planning energy systems that consider different sectors in the sense of multi-energy system planning and the planning of transmission and distribution systems, such as \cite{PyPSA, Schwaeppe2022}. However, since these methods require extensive computational resources, it is necessary to simplify the operation of individual sectors and the infrastructure to determine an optimized solution. Moreover, during simplification, the distribution grids were often represented as passive grid nodes, and the power flows, which are transmitted via the distribution grid infrastructure, were neglected.
For the representation of operational constraints in the electric grids, the \ac{FOR} has proven applicability \cite{Fortenbacher2019}. In previous work, \cite{boettcher2023_for} has described the use of \ac{FOR} to represent distribution grid constraints in power system planning approaches. However, the relationship between distribution grid expansion and the associated systemic costs was represented in a simplified way. 
With this paper, we present a method to create a solution space, the \ac{FPR}, for the representation of distribution grids by modeling the operative boundaries of the planned infrastructure combined with the necessary investment costs. In this way, we want to enable the operational limits and the required infrastructure investments of active distribution grids to be accurately represented in systemic planning approaches. 

\subsection{State of the art}
Large-scale system studies are currently examining the future evolution of the energy system to identify the most cost-effective paths. However, most of these studies focus on the transmission system and overlook the potential congestion within the distribution system below. To address the representation of the physical distribution grid constraints in multi-energy system optimization modeling, we proposed an approach in \cite{boettcher2023_for}, in which we incorporated the concept of the \ac{FOR} to represent grid constraints in system studies without significantly increasing computational complexity. In this paper, we further present the consideration of an additional dimension so that different grid expansion stages at the distribution grid level can be considered in energy system planning.

\subsubsection{Power system planning}
There are various methods available for evaluating and planning future energy systems. For example, \cite{PyPSAEur} provides an open optimization framework for European transmission system planning, and \cite{Schwaeppe2022} offers a model to explore alternative paths in intersectoral technology planning. However, these approaches are primarily concerned with determining energy system configurations and do not dive into the detailed representation of electrical distribution grids.
Nevertheless, an integrated formulation of transmission and distribution system expansion planning can be found in the literature \cite{Muller2019}. A comprehensive overview is also presented in \cite{Delgado2021}. However, these formulations were formulated for a limited spatial scope, e.g., small regions lacking suitability for large-scale system studies. Exemplary more extensive studies are, for example, \cite{Muller2019, Baecker2022} focusing on Germany. However, the physical constraints of the distribution grids are not represented in the model formulation. In \cite{Baecker2022}, only the physical properties for low-voltage grids are considered with the outlook to consider multiple voltage levels. The approach in \cite{Allard2020} does a global assessment of the European power system for 2050, considering distribution grids, their categorization into rural, suburban, and urban groups, and incorporating grid constraints with linearized power flows at specific TSO grid nodes. However, it is not widely applied in large-scale system studies. 
While distribution grids are considered in some energy system planning approaches, additional expansion stages and respective physical changes to the integrable power to the distribution grid level are not modeled in detail.

\subsubsection{Feasible operation region}
The \ac{FOR} defines safe operating conditions for the distribution grid, containing all valid interconnection power flows, and it serves as a means to determine the limits on power exchange between the upstream transmission and downstream distribution systems, also referred to as \ac{IPF} \cite{Gotzens2019}. These limits are influenced by various factors such as grid infrastructure, including lines and transformers, as well as loads and generators, with the problem formulation relying on power flow equations. 
In recent years, extensive research has been conducted to establish operational limits for \ac{DSO} to provide services to the upstream system, such as ancillary services or market schemes. In \cite{FORTENBACHER2020105668}, a concept for deriving operational limits as a \ac{FOR} of the distribution system is presented.

\subsection{Contributions of this work}
In this research paper, we propose an enhanced method for power system planning. Our approach includes the cost of expanding distribution grids and considers physical limitations for a more comprehensive analysis. We achieve this by designing a new solution space that considers both the hosting capacity and the cost of different statuses of distribution grid expansion. Building on the \ac{FOR}, we present the \ac{FPR}, which thus represents an aggregate representation of the costs and integrable power into the distribution grids. We, therefore, extend the procedure already presented in \cite{boettcher2023_for} by using the FPR.


%% file: 02_methodology.tex
\section{Methodology}
\label{sec: methodology}
In the following, the method of creating \ac{FPR} is described. The \ac{FPR} is a novel aggregation method for \acl{DG}s. It represents \acl{DG}s by calculating their feasible \acp{IPF} at the substation, reflecting flexibility potentials and technical limitations (\ref{subsec: FORgen}). Moreover, it represents the grid cost of the given \acl{DG}, comprising multiple different expansion levels of the grid (\ref{subsec: caesar}). Thus, distribution grids are represented in an aggregated manner by maintaining the physical information for integrated transmission and distribution grid planning.

\begin{figure}[ht]
     \centering
     \includegraphics[width=.88\linewidth]{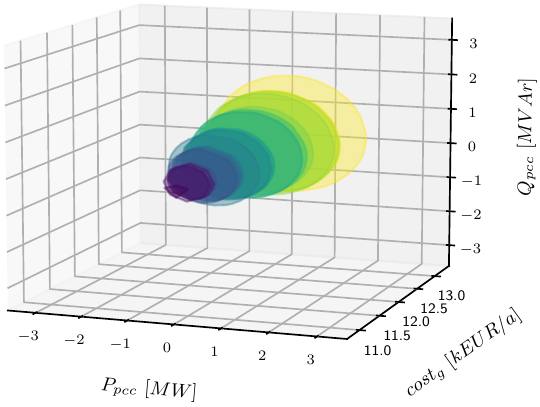}
     \caption{Exemplary Feasible Planning Region (FPR)}
     \label{fig: exemplary fpr}
\end{figure}

Fig. \ref{fig: exemplary fpr} shows an exemplary \ac{FPR}. Here, a three-dimensional space includes an active $P_{pcc}$ and reactive $Q_{pcc}$ power-axis representing the \ac{IPF} at the point of common coupling (${pcc}$), and a cost-axis representing the grid cost $cost_{g}$ at different expansion stages $g \in \mathcal{G}$. The \ac{FPR} is located within this space, consisting of a set of \acp{FOR}, aligned along the cost-axis. Each \ac{FOR} represents a grid's feasible interconnection power flow, see \ref{subsec: FORgen}. 
In the presented example the lower cost \acp{FOR} represent a lower degree of freedom for potential \acp{IPF}. With higher expansion levels of the grid and accompanying higher costs, the degree of freedom of possible \acp{IPF} also increases. Thus, the \ac{FPR} represents the relation between the \ac{FOR} and the respective grid cost, which is visualized in this example. 

\begin{figure}[ht]
	\centering
	\input{figures/FPR_workflow.tex}
	\caption{Workflow to obtain a Feasible Planning Region}
	\label{fig: FPR workflow}
\end{figure}
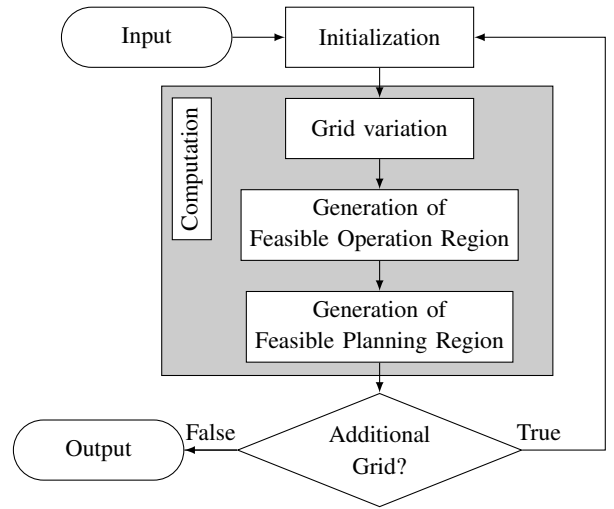

Fig. \ref{fig: FPR workflow} presents the workflow for generating the \ac{FPR} for a multi-voltage-level \acl{DG}.

This workflow is organized into several processing steps for pre-processing and calculations. It is discussed by first describing the necessary input of the workflow in \ref{subsec: input data}, followed by the computation processes in \ref{subsec: caesar} describing the grid variation process, \ref{subsec: FORgen} presenting the generation of the \ac{FOR},  \ref{subsec: FPRgen} describing how the \ac{FPR} is set up, and concluding the consideration of multiple-voltage-levels in \ref{subsec: initialization}.

\subsection{Input Data}
\label{subsec: input data}
In this section, we outline the essential input data requirements for the workflow. These inputs serve as the foundation for subsequent computations and are crucial for accurate results. 
To adequately represent distribution grids in power system modelling, representative grid models are needed for the approach. There are several procedures discussing, how representative grid models can be determined for adequate extrapolation for a systemic result \cite{boettcher2023_extrapolation}. In addition to the representative grid models, generator and load models, as well as the associated profiles and parameterizations, are required as input.
Initially, the grid models are simplified to a single voltage level due to the voltage-specific nature of the grid expansion measures outlined in \ref{subsec: caesar}. This consolidation ensures compatibility with the grid expansions described in that section. The representative distribution grids are furthermore categorized based on their urbanization levels, which capture distinct spatial characteristics. 
These input distribution grid models and their interconnection into one multi-voltage-level distribution grid, are the basis for constructing the \ac{FPR}.

\subsection{Grid Variation}
\label{subsec: caesar}
With this process, a variety of distribution grid expansion stages is generated. Based on the variation of the supply task on the initial grid model, congested grids are generated, which are subsequently modeled to congestion-free grids by grid expansion measures. The generation process is based on \cite{boettcher2023_cost_drivers}. The process can be delineated into three sequential steps, the \textit{generation of supply task scenarios}, \textit{grid expansion}, and \textit{grid cost determination}.
As a result, a set of new distribution grids with several expansion stages and their respective grid cost is generated.

\subsubsection{Generation of Supply Task Scenarios} 
By means of this process step, the probability of different future supply tasks on a distribution grid is mapped. A supply task scenario defines the supply task by changing distribution grid parameters. 
Thus, based on the initial distribution and total capacity of load and generation, a randomly changed supply task is created on the same grid topology. To ensure that different distributions and individual load and generation capacities, as well as types, per node occur, the allocation process is repeated, similar to a Monte Carlo simulation. 
In order to generate further expansion stages in addition to the expansion stages through random distribution, the generation and load capacities are also increased sequentially. 
These distribution grid scenarios are then used to determine the necessary grid expansion. 

\subsubsection{Grid Expansion}
Based on the scenarios generated, the resulting grid utilization is then determined by means of power flow calculation. If grid congestions occur, grid expansion measures are carried out on the grid model in order to eliminate the congestion. 
Based on the heuristic approaches and planning principles of DSOs, planning-relevant grid use cases are defined based on the supply task. These include high load and high feed-in cases, which represent worst-case power flows on the grid infrastructure. By applying the grid use cases to the grid models, thermal limits or voltage band violations can be detected.
In this method, grid expansion is performed with a heuristic method to ensure reliable grid operation. First, the elimination of thermal limit violations is done with reinforcement measures, by replacing the violated grid assets with equipment capable of withstanding higher thermal loads or installing additional parallel assets. Afterwards, the elimination of voltage band violations is performed, by performing line separations at two-thirds of the affected line. The grid expansion results in a set of new \acl{DG}s, with different reinforcement measures applied.

\subsubsection{Grid Cost Determination}
Within this step, the grid cost for each distribution grid is computed. The total grid cost of each distribution grid is obtained by summing up the total line and transformer cost of each element, considering material and installation costs. This results in the formulation 
\begin{equation}
    c_{total} = \sum_{i = 0}^{N} (c_{i, install} + c_{i, mat}) \cdot l_{line} + \sum_{j=0}^{M} c_{j,trafo} 
\end{equation}
where $c_{i, install}$ describes the installation cost per line length for lines, based on the spatial urbanization, $c_{i, mat}$ is the material cost per line length for the given line type, $c_{trafo}$ is the transformer cost and $l_{line}$ is the line length. Moreover, $N$ is the number of lines, and $M$ is the number of transformers within the grid.

In conclusion, the grid expansion results in a set of \acl{DG} expansion stages, one for each supply task scenario. These are based on the initial \acl{DG} but with expansion measures applied accordingly to the supply task scenarios. Due to the grid reinforcement measures, new interconnection power flows and grid costs occur for each \acl{DG}.

\subsection{Generation of Feasible Operation Region}
\label{subsec: FORgen}
Subsequently, this process step receives a set of \acl{DG} expansion stages and generates one \ac{FOR} for each of these \acl{DG}s. 
The \ac{FOR} is commonly known for analyzing synchronous generators. It describes an area within the two-dimensional active- and reactive-power plane, visualizing all feasible operation points and, therefore, feasible \acp{IPF} through the point of common coupling (${pcc}$). In recent research, the \ac{FOR} is also part of the \ac{TSO} and \ac{DSO} coordination concepts, representing all feasible \acp{IPF} through the substation of the given distribution grid. Here, it is used as a communication method for the \ac{DSO} to communicate its flexibility potential to the \ac{TSO} \cite{Papazoglou2022}. 
Within this work, the \ac{FOR} is used for reducing the distribution grid complexity while maintaining information about its physical infrastructure and components \cite{boettcher2023_for}. Here, the distribution grid is reduced to one node, while the information about grid constraints and flexibility potentials are maintained by the feasible \acp{IPF}, represented by the \ac{FOR} itself. 

To compute the FOR, an optimization-based approach is chosen, performing an adapted non-linear optimal power flow on the distribution grid, following the general formulation 
\begin{equation}
    \begin{aligned}
        & \min \limits_{y}\ f(x, y) \\
        & \mathrm{s.\ t.} \ \ c_{i}^{\mathrm{E}}(x, y) = 0 \quad \forall i = 1, \dots, n \\
        & \quad \; \; \; \; \; c_{j}^{\mathrm{I}}(x, y) \leq 0 \quad \forall j = 1, \dots, m , \\
    \end{aligned}
    \label{opt}
\end{equation}
where $f(x,y)$ describes the objective for optimizing the interconnection power flow at the substation, $c_i^E(x,y)$ defines equality constraints, which are typically related to physical relation, and $c_j^I(x,y)$ defines inequality constraints, typically related to technical limitations \cite{boettcher2023_for}. The exact formulation of this optimization problem and its iterative process is based on the algorithm described in \cite{Lopez2021}. This calculation process results in a \ac{FOR} related to the given distribution grid and its cost. Since a complete set of \acl{DG} expansion stages is computed with the presented algorithm, a set of \acp{FOR} is the outcome.

\subsection{Generation of Feasible Planning Region}
\label{subsec: FPRgen}
This section describes the process for constructing a \ac{FPR} out of the set of \acp{FOR}. The \ac{FPR} is generated by aligning the resulting \acp{FOR} along the cost axis. This cost axis represents the grid cost for the given distribution grid variation. As a result, the \ac{FPR} illustrates all feasible \acp{IPF} for a given distribution grid and its expansion levels, within a three-dimensional plane (cf. Fig. \ref{fig: exemplary fpr}).

Within the generation process of \acp{FPR}, conflicts occur when different grid expansion stages have the same grid cost related to them. In this case, multiple \acp{FOR} are located at the same cost-axis position. To repair these conflicts, the covered area of the conflicting \acp{FOR}, representing the feasible \ac{IPF}, are compared, and the larger one is selected. As a result, the \ac{FOR}, which offers higher flexibility, is chosen. The resulting discreet solution space thus describes the relationship between possible representable load flows and the associated costs of each grid expansion state. 

\subsubsection{Multi-Voltage-Level Feasible Planning Region}
\label{subsec: initialization}
Since the representation of distribution grids in power system planning requires the representation of multiple voltage levels, a bottom-up approach to modeling multiple voltage levels was developed.


The bottom-up calculation approach performs the \ac{FPR} calculation of low- to higher-voltage distribution grids. Initially, the computation focuses on low-voltage \acl{DG}s, with the supply task consisting solely of connected load and generation units, enabling the computation of the \ac{FPR} using the prescribed computation steps. Subsequently, medium-voltage-level distribution grids are computed, incorporating load, generation, and the already-calculated low-voltage \acp{FPR}. For performing the \ac{FPR} calculation process, the already calculated low-voltage level \acp{FPR} are introduced into the medium voltage grids. The \acp{FPR} are approximated by generation units covering the area of the FOR and are expandable according to the \ac{FPR}. The \acp{FPR} are approximated by generation units covering the area of the FOR and are expandable according to the \ac{FPR}. This approach facilitates the computation of \acp{FPR} across multiple voltage levels.

Within the workflow, the bottom-up modelling is achieved using a loop with the initialization process. The loop allows the successive computation of distribution grids from low to higher voltage level grids (cf. Fig. \ref{fig: FPR workflow})

%% file: figures/FPR_workflow.tex
\begin{tikzpicture}[node distance=0.4cm and 0.7cm]     
	\node[draw,
	rounded rectangle, 
	minimum width=2.5cm,
	minimum height=0.8cm,
	align=left] 
	(start) 
	{\small Input};

 	\node[draw,
	rectangle, 
	minimum width=2.5cm,
	minimum height=0.8cm,
	right= of start,
	align=center] 
	(initialization) 
	{\small Initialization};

    \draw[black, fill=black!20] (0.2,-0.65) rectangle (5.4,-4.5);

 \filldraw[black] (0.6, -0.8)  node[anchor=east, rotate= 90, text=black, draw, fill=white]{\small Computation};
    
	\node[draw,
	rectangle, 
    fill=white,
	minimum width=2.5cm,
	minimum height=0.8cm,
	below= of initialization,
	align=center] 
	(caesar) 
	{\small Grid variation};
	
	\node[draw,
	rectangle, 
    fill=white,
	minimum width=2.5cm,
	minimum height=0.8cm,
	below= of caesar,
	align=center] 
	(FOR) 
	{\small Generation of \\
    \small Feasible Operation Region};

    \node[draw,
	rectangle, 
    fill=white,
	minimum width=2.5cm,
	minimum height=0.8cm,
	below= of FOR,
	align=center] 
	(FPR) 
	{\small Generation of \\
    \small Feasible Planning Region};
	
	\node[draw,
	diamond, 
	minimum width=1cm,
	minimum height=0.8cm,
	aspect=2.5,
	below= of FPR,
	align=center] 
	(decision) 
	{ \small Additional\\
    \small Grid?};
	
	\node[draw,
	rounded rectangle, 
	minimum width=2.5cm,
	minimum height=0.8cm,
	left= of decision,
	align=center] 
	(end) 
	{\small Output};

	\draw[-latex] 
	(start) edge (initialization)
    (initialization) edge (caesar)
	(caesar) edge (FOR)
    (FOR) edge (FPR)
	(FPR) edge (decision)
    (decision) edge (end);
	
	\draw[-latex]
	(decision) -- (end)
	node[pos=0.5, above]{\small False}
	(decision) -- +(3,0) node[pos=0.2, above]{\small True} |- (initialization);

\end{tikzpicture}

%% file: 02_modelling.tex
\section{Modeling}
\label{sec: modeling}

The \ac{FPR} has been developed to be used in power system planning. Therefore, this section presents the integration of the \ac{FPR} into a multi-energy-system (MES) model for performing a linear generation, storage, and transmission capacity expansion problem (CEP). 
This section, therefore, first presents a general capacity expansion problem formulation, which is defined in \ref{subsec: capacity expansion problem} to determine the necessary requirements for the formulation of the solution space of the \ac{FPR}. Additionally, the solution space of the \ac{FPR} is modeled accordingly in \ref{subsec: linearization FPR}. Concluding, the \ac{FPR} is integrated into the capacity expansion problem (\ref{subsec: integration MES}).

\subsection{Capacity Expansion Problem}
\label{subsec: capacity expansion problem}
Capacity expansion problems are needed to assist transmission system operators, and power plant operators in optimizing their investment decisions. The objective is to optimize investment decisions based on infrastructure restrictions, capital costs, and operational costs. The general and concise formulation of a \ac{MES} capacity expansion problem is
\begin{alignat}{3}
\min_{x \in \mathbb{R}^d}       &\quad& C(x)  &&     &                           \\
\text{s.t.: }                   &\quad& f_i(x) && \leq b_i  &\quad \forall i=1,\dots, N
\end{alignat}
where $C(x)$ is the objective function, describing all capital and operation costs of the MES, and $f_i(x)$ is a set of constraint functions, describing technical limitations and physical relations \cite{boettcher2023_for}.

The formulation of a capacity expansion problem is commonly non-linear \cite{Conejo2016}. This is because, based on the many possible possibilities in large areas of consideration, such as entire countries, the solution space becomes very large, and thus, the computation time can be reduced to a reasonable time by linearizing the problem. Since the problem size of the capacity expansion problem is large-scale and linear approximations are valid when considering transmission systems, a linear problem formulation is also chosen for the integration of the \ac{FPR}. Therefore, the objective function and the constraints are formulated as linear functions, formulated as
\begin{align}
    C(x)=\sum_{j=1}^d (C_{j,1} x_j + C_{j,0}) \\
    f_i(x)=\alpha_i^T x + \beta_i \quad \alpha, \beta \in \mathbb{R}^d \quad \forall i=1,\dots, N
\end{align}
This results in increasing computational performance and more straightforward integration of the \ac{FPR} in the given MES model. 

\subsection{Linearization of the Feasible Planning Region}
\label{subsec: linearization FPR}
Since a linear formulation of the capacity expansion problem is used, the solution space of the \ac{FPR} must be modelled accordingly to be integrated into the \ac{MES} model. The aim of this linearization process is to model a continuous representation of discreet expansion stages within the \ac{FPR}. Based on this simplification, it is assumed that discrete expansion measures are no longer represented and partial expansion can be performed. With a view to a systemic consideration, it is assumed that this influence does not have a significant impact on the overall result.

Regression is performed for the linearization of the \ac{FPR}, resulting in a convex solution space for the \ac{MES} formulation. Based on the regression, a linear relationship between active power $P_{pcc}$ and reactive power $Q_{pcc}$  and the corresponding grid costs $c_g$  can be presented. The information provided by the \ac{FPR} also allows a non-convex representation of the solution space for other model formulations. This is the focus of future research.

\subsection{Integration into a Multi-Energy-System Model}
\label{subsec: integration MES}

In multi-energy system models the representation of distribution grids are can be modeled as separate DSO nodes which are connected to the TSO nodes (TSO/DSO-link). Such multi-energy system model with capacity expansion problem formulation must therefore be expanded for the consideration of \acl{DG}s (\acp{FPR}). This work uses a similar approach as described in \cite{boettcher2023_for}. 

It should be noted, that in a power system planning process, the selection of representative distribution grids depends on the cluster method and the selection of representative distribution grid regions. If a representative region has been identified, a representative grid model can either be modeled for it or assigned from existing grid models. The cluster method is also used to determine which parameters can be used to extrapolate from a representative grid model to the DSO node to be mapped within a planning framework. This problem is discussed in detail in \cite{boettcher2023_for} and \cite{boettcher2023_extrapolation}.

The objective function of the capacity expansion problem is modified to include the cost associated with \acl{DG}s. This modification is expressed as
\begin{equation}
    \min_{x \in \mathbb{R}^d} C(x) + \sum_{n} c_{n,d} M_{n,d} + \sum_{n} o_{n,d} f^{pcc}_{n,d,t} 
    \label{capitalcostDSO}
\end{equation}
where $C(x)$ is the objective function of the previously described capacity expansion problem, $M_{n,d}$ is the power capacity integrated into the distribution grids, $f^{pcc}_{n,d,t}$ the power flow of the TSO/DSO interconnection point, $c_{n,d}$ represents the capital costs ($c_{g}$) of the distribution grids and $o_{n,d}$ the operational costs of the distribution grids. Moreover, $d \in \mathcal{D}$ describes the different \acl{DG} classes, representing the urbanization level of the \acl{DG}. $n \in \mathcal{N}$ represents the TSO/DSO interconnection points. As a result, the linear \ac{FPR} can be integrated through the additional terms, which allows the representation of different expansion levels of \acl{DG}s.

Furthermore, two additional constraints are defined to limit the \ac{IPF} between the \acl{DG} and the \acl{TG}.
\begin{align}
    {M}_{n,d,min} \leq M_{n,d} \leq {M}_{n,d,max} \label{linklimitDSO} \\
    {f}^{pcc}_{n,d,t,min} M_{n} \leq f^{pcc}_{n,d,t} \leq  {f}^{pcc}_{n,d,t,max} M_{n,d} \label{linklimitDSO1}
\end{align}
Here, the interconnection power capacity $M_{n,d}$ can be expanded as needed, according to the given power capacity limits $M_{n,d,min}$ and $M_{n,d,max}$. The power capacity sets the boundaries for the interconnection power flow $f^{pcc}_{n,d,t}$ between the \acl{DG} and the \acl{TG}, where $f^{pcc}_{n,d,t,min}$ and $f^{pcc}_{n,d,t,max}$ are adapting the unit, sign, and are scaling the power capacity accordingly.

Hence, the adapted objective function and the additional constraints allow the integration of the linear \ac{FPR} into the existing capacity expansion problem.

%% file: 03_results.tex
\section{Results}
In the following, we present the form of the solution space for several grid topologies and discuss the implications of structural differences considering the degree of urbanization and voltage level. Furthermore, we showcase the applicability of \ac{FPR} in an open-source model for system planning.

To model the results, publicly available grid models have been used \cite{Meinecke2020}. The grid models here are already divided according to different voltage levels and degrees of urbanization. This division was adopted in the generation of these results. Since a large number of scenarios are determined for the preparation of the FPR (distribution and dimensioning of generation and load), only a limited number of grid expansion stages are presented below for presentation purposes.

\subsection{Comparison of different Feasible Planning Regions}
\label{ch:results_comp_FPR}
To identify various Feasible Planning Regions (FPRs), the model grids are processed using the previously explained procedure.

\subsubsection{Voltage Level}
\label{ch:results_voltage_level}
To compare different \acp{FPR}, exemplary results for the low-, medium- and high-voltage levels are presented below. 

Fig. \ref{fig: LV FPR} shows the \ac{FPR} of a low voltage grid. 

\begin{figure}[ht]
     \centering
     \includegraphics[width=.88\linewidth]{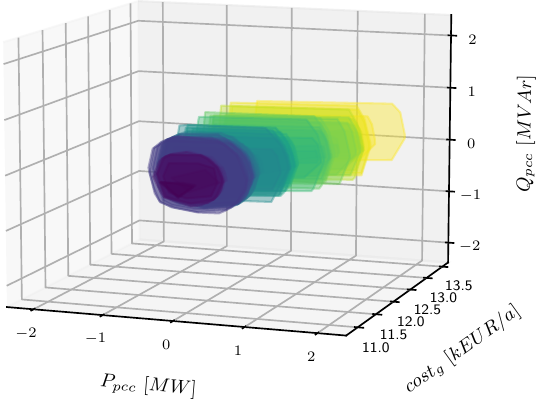}
     \caption{Low Voltage Feasible Planning Region (FPR)}
     \label{fig: LV FPR}
\end{figure}
 The low voltage grid results show that as grid costs increase, the \ac{FOR} in the \ac{FPR} increases. Based on the expansion of the resources, a larger \ac{IPF} over the point of common coupling is therefore possible, as expected. Moreover, the shape of the \ac{FOR} does not change continuously with the expansion of the grids. This is because the distribution of operating resources occurs statistically and not optimally in the individual grid expansion stages. Moreover, in this low-voltage grid, it can be seen that the reactive power $Q_{pcc}$ is more constrained than the active power $P_{pcc}$. Since a rural low voltage grid was used in this example and there are higher line lengths, the voltage bands are the limiting restrictions, which affects the reactive power provision at the point of common coupling. 

\begin{figure}[ht]
     \centering
     \includegraphics[width=.88\linewidth]{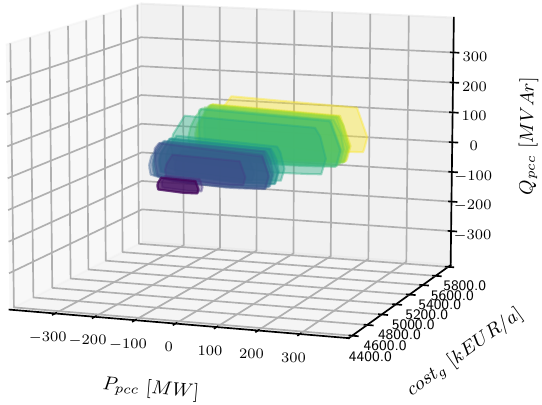}
     \caption{Medium Voltage Feasible Planning Region}
     \label{fig: MV FPR}
\end{figure}

Fig. \ref{fig: MV FPR} presents the result of a multi-voltage level \ac{FPR}. 

In this example, the bottom-up approach described above has been used to determine \acp{FPR} across multiple voltage levels. Again, a rural grid was used for this result. In the shape of the \ac{FPR}, a gradual increase in \acp{FOR} can be observed. Similar to the low voltage grid shown above, the limitation of the reactive power $Q_{pcc}$ at the point of common coupling can also be observed here. The active power $P_{pcc}$ is almost symmetrical in each expansion stage's positive and negative directions. 

\begin{figure}[ht]
     \centering
     \includegraphics[width=.88\linewidth]{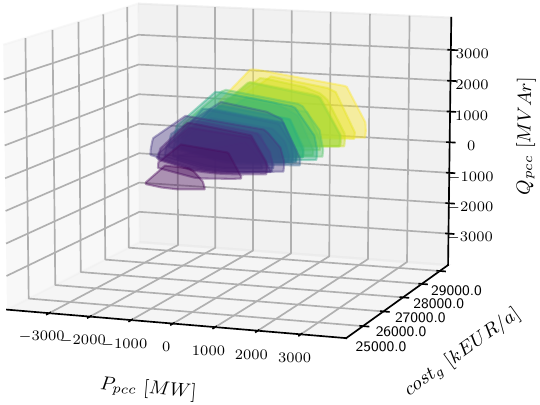}
     \caption{High Voltage Feasible Planning Region}
     \label{fig:HV FPR}
\end{figure}

In Fig. \ref{fig:HV FPR}, an exemplary \ac{FPR} for the high-voltage level is shown. 

A high-voltage grid with mixed urban and rural structures was used to compute the \ac{FPR} for the high-voltage level. It can be seen that the power transmission is significantly higher than in the medium and low voltage grids. In addition, the shape of the individual \acp{FOR} of the different grid expansion stages is less symmetrical due to the mixed grid structure. It is also noticeable that the individual grid expansion stages' FOR does not increase continuously. This is due to the meshed grid topology of the high-voltage grid used here. As soon as a power limit is not met at a substation, an expansion is necessary, which strongly depends on the previously generated distributions of generation and load in the scenarios. Thus, a different distribution of the same loads and generation may result in lower grid costs, which is reflected in the FPR. In such a case, linearization partially underestimates the grid costs in the high voltage when it is used in power system planning.

\subsection{Examplary Study Case}
\label{ch:results_study_case}
To present the application of \ac{FPR} and the effects on the planning result in an exemplary way, an exemplary use case is considered in which two scenarios are compared. Subsequently, the results are discussed. For modelling the European multi-energy system, PyPSA-Eur-Sec is used \cite{PyPSAEurSec}. PyPSAEur-Sec is an open-source, cross-border and cross-sector energy system model capable of calculating the transmission grid of the complete ENTSO-E area. It solves a capacity expansion problem, including expansion measures of the transmission system, generation units and storage facilities.
This study case focuses on a reduced European energy system with 37 nodes, which results in one node per country. The investigation year 2030 is selected. The weather data are based on historical data and according to 2013. Moreover, a green field approach is selected, except for already installed conventional power plants like coal, lignite, oil and nuclear.

\subsubsection{Scenarios}
\label{ch:results_scenarios}

This paper explores two scenarios for evaluating the application of the \ac{FPR}. Both scenarios are identical, differing in that the first one disregards the application of the \ac{FPR} and thus neglects the distribution grid costs in the optimization process. In this case, the links connecting the TSO and the DSO node incur no investment costs and have unlimited power transmission capacity. Conversely, the second scenario accounts for distribution grid costs, which are factored into the definition of \ac{FPR}. Expanding the distribution grids in this scenario involves investment costs associated with this defined planning region.

\subsubsection{Study Case Results}
\label{ch:results_PyPSA}
\begin{figure}[h]
     \centering
     \includegraphics[width=.98\linewidth]{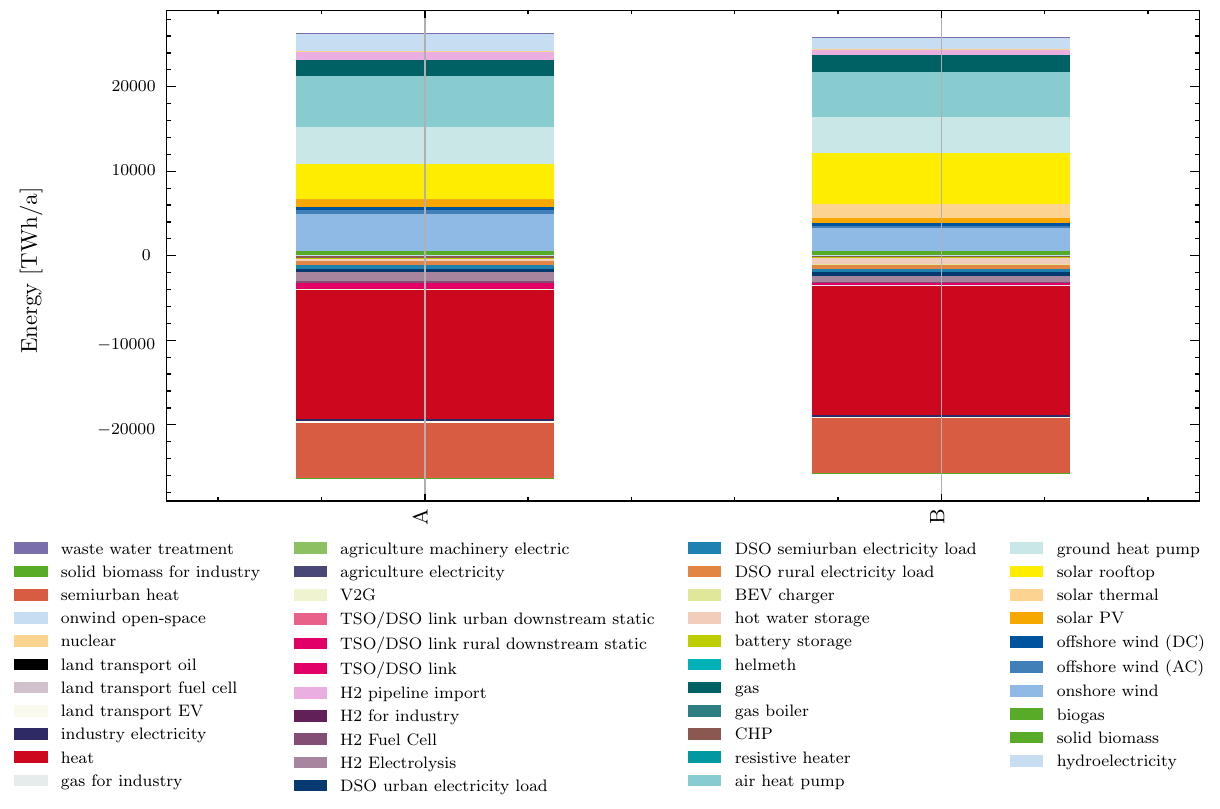}
     \caption{Energy Balances of Study Case}
     \label{fig:pypsa}
\end{figure}

The results of the optimization of the Study Case scenarios show an overall result for the European Energy System.
Fig. \ref{fig:pypsa} illustrates sector-dependent energy generation and demand. 

The ordinate of the two-dimensional plane presents the provided and consumed energy over the year 2030, where positive values denote the provision of energy and negative values its consumption. The abscissa lists the different scenarios, where scenario A contains no distribution grid cost and scenario B includes distribution grid cost. The colour and size of the given bar plot segments indicate the contribution of different technologies and sectors. Comparing both scenarios in Fig. \ref{fig:pypsa}, a deviation can be seen e.g. in energy provision via solar rooftop and wind turbine power plans. The integration of distribution grid cost results in a higher expansion of solar rooftop generation units. The generation via photovoltaic rooftop plants is located at the distribution level and near the energy demand. Power generation near the demand locations reduces transmission losses and possibly avoids expansion measures of the distribution grids when taking the \ac{FPR} into account. In scenario A, the missing solar rooftop generation is compensated by wind turbines.
On the demand side of Fig. \ref{fig:pypsa}, reduced losses of the distribution grid link are illustrated. Denoted as TSO/DSO link, the transmission losses between the transmission and distribution grid are visualized. A loss decrease in scenario B is visible, supporting the assumption that increasing solar rooftop generation reduces the interconnection power flow.

Based on the results presented here, it can be deduced that the consideration of pyhsical constraints as well as the costs of the distribution grids in the optimization can lead to different results in the technology selection and thus the optimal system configuration. Since the results here are exemplary in nature to show the applicability of the\ac{FPR}, a detailed investigation of the influence on power system planning results is pending, which is the objective of future research.

%% file: 04_conclusion.tex
\section{Conclusion}
Using the method presented in this paper and the new method of FPR, it is now possible to take into account different levels of expansion and, respectively, the costs of the electrical distribution grid infrastructure and their physical limits in energy system planning models. The results show the influence of different levels of urbanization and voltage levels on the shape of the FPR. Furthermore, the influence of FPR on a power system planning result was shown as an exemplary study case. In further research, the influence of FPR on energy system planning results will be further explored and the method will be applied to representative grids.